\documentclass[twocolumn,superscriptaddress,showpacs,preprintnumbers,amsmath,amssymb,prx,aps]{revtex4-1}

\usepackage{amsmath}
\usepackage{graphicx}
\usepackage{dcolumn}
\usepackage{bm}
\usepackage{color}
\usepackage[colorlinks,linkcolor=blue,urlcolor=blue,citecolor=blue]{hyperref}
\usepackage{epstopdf}

\begin{document}

\newcommand{\ie}{{\it i.e.}}
\newcommand{\eg}{{\it e.g.}}
\newcommand{\etal}{{\it et al.}}


\title{Nodal superconductivity and superconducting dome in new layered superconductor Ta$_4$Pd$_3$Te$_{16}$}%

\author{J. Pan}%
\affiliation{State Key Laboratory of Surface Physics, Department of Physics, and Laboratory of Advanced Materials, Fudan University, Shanghai 200433, China}

\author{W. H. Jiao}%
\affiliation{Department of Physics, Zhejiang University, Hangzhou 310027, China}

\author{X. C. Hong}\author{Z. Zhang}\author{L. P. He}\author{P. L. Cai}\author{J. Zhang}%
\affiliation{State Key Laboratory of Surface Physics, Department of Physics, and Laboratory of Advanced Materials, Fudan University, Shanghai 200433, China}

\author{G. H. Cao}%
\affiliation{Department of Physics, Zhejiang University, Hangzhou 310027, China}
\affiliation{Collaborative Innovation Center of Advanced Microstructures, Nanjing University, Nanjing 210093, P. R. China}

\author{S. Y. Li}%
\email{shiyan\_li@fudan.edu.cn}
\affiliation{State Key Laboratory of Surface Physics, Department of Physics, and Laboratory of Advanced Materials, Fudan University, Shanghai 200433, China}
\affiliation{Collaborative Innovation Center of Advanced Microstructures, Fudan University, Shanghai 200433, P. R. China}

\date{\today}

\begin{abstract}
We measured the low-temperature thermal conductivity of a new layered superconductor with quasi-one-dimensional characteristics, the ternary telluride Ta$_4$Pd$_3$Te$_{16}$ with transition temperature $T_c \approx$ 4.3 K. The significant residual linear term of thermal conductivity in zero magnetic field and its rapid field dependence provide evidences for nodes in the superconducting gap. By measuring resistivity under pressures, we reveal a superconducting dome in the temperature-pressure phase diagram. The existence of gap nodes and superconducting dome suggests unconventional superconductivity in Ta$_4$Pd$_3$Te$_{16}$, which may relate to a charge-density wave instability in this low-dimensional compound.
\end{abstract}


\maketitle

\section{Introduction}
Finding unconventional superconductors and understanding their superconducting mechanism is one of the main themes in condensed matter physics \cite{Norman-review}. The term ``unconventional" firstly means the superconducting pairing mechanism is not phonon-mediated. This usually manifests as a superconducting dome neighbouring a magnetic order in the phase diagram, and spin fluctuations are considered as the major pairing glue \cite{Norman-review}. Secondly, the term ``unconventional" means the wave function of Cooper pairs is not $s$-wave. Symmetry imposed nodes (gap zeros) are often observed, such as in $d$-wave cuprate superconductors and heavy-fermion superconductor CeCoIn$_5$ \cite{Tsuei-review,CeCoIndwave}, and in $p$-wave superconductor Sr$_2$RuO$_4$ \cite{SrRuO-review}. Note that the iron-based superconductors are exceptions, likely with the form of $s_\pm$-wave \cite{Fe-review}. The superconducting gap symmetry and structure provide important clue to the underlying pairing mechanism.

Unconventional superconductivity usually resides in quasi-two-dimensional (Q2D) compounds, such as cuprate and iron-based superconductors, CeCoIn$_5$, Sr$_2$RuO$_4$, and organic superconductors $\kappa$-(BEDT-TTF)$_2$X \cite{Norman-review}. When further reducing the dimensionality, namely in quasi-one-dimensional (Q1D) superconductors represented by the organic compounds (TMTSF)$_2$X (X = PF$_6$, ClO$_4$) \cite{Jerome,Bechgaard}, the pairing symmetry and mechanism are also likely unconventional \cite{Zhang-review}. In this sense, low dimensionality is important for the appearance of unconventional superconductivity \cite{IIMazin}.

\begin{figure} [t]
\centering
\includegraphics[clip,width=8.5cm]{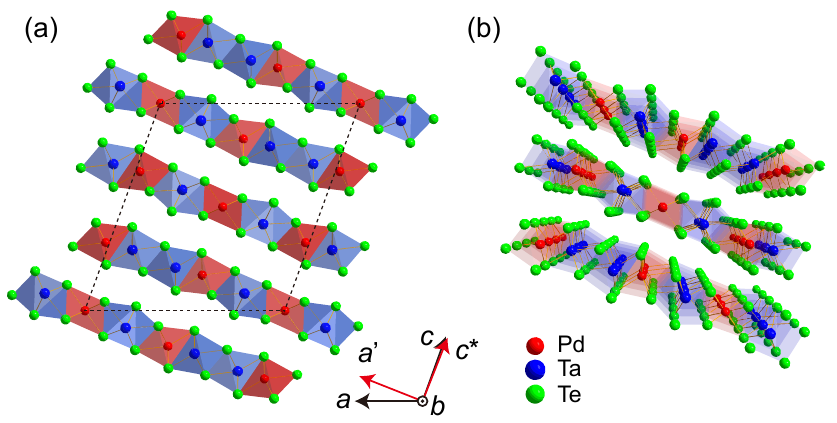}
\caption{\label{Crystal}
Crystal structure of Ta$_4$Pd$_3$Te$_{16}$. (a) A view parallel to the $ac$ plane. The compound crystallizes in space group $I$2/m with a monoclinic unit cell of $a$ = 17.687(4) {\AA}, $b$ = 3.735(1) {\AA}, $c$ = 19.510(4) {\AA}, and $\beta$ = 110.42$^\circ$. The crystal structure has relatively flat Ta$-$Pd$-$Te layers in the ($\overline{1}$03) plane, which is the largest natural surface of as-grown single crystals. For convenience, we define $a'$ direction as [301], so that the ($\overline{1}$03) plane is the $a'b$ plane. $c^*$ is the direction perpendicular to the $a'b$ plane. The Pd atoms are octahedrally coordinated, forming edge-sharing PdTe$_2$ chains along the $b$ axis. (b) A three-dimensional perspective view along the $b$ axis. The PdTe$_2$ chains are separated by TaTe$_3$ chains and Ta$_2$Te$_4$ double chains.}
\end{figure}

Recently, the ternary telluride Ta$_4$Pd$_3$Te$_{16}$ \cite{Mar} was found to be a new layered superconductor with Q1D characteristics \cite{TPT-Jacs}.  The $T_c$ is about 4.5 K at ambient pressure. It has relatively flat Ta$-$Pd$-$Te layers in the ($\overline{1}$03) plane, which contains PdTe$_2$, TaTe$_3$, and Ta$_2$Te$_4$ chains along the crystallographic $b$ axis, as illustrated in Fig.~\ref{Crystal}. It will be very interesting to check whether there exists unconventional superconductivity in this low-dimensional compound.

In this paper, we present the low-temperature thermal conductivity measurements of Ta$_4$Pd$_3$Te$_{16}$ single crystal down to 80 mK, which clearly demonstrates that there are nodes in the superconducting gap. Furthermore, a superconducting dome in the temperature-pressure phase diagram is revealed by resistivity measurement under pressures up to 21.9 kbar. These results suggest unconventional superconductivity in Ta$_4$Pd$_3$Te$_{16}$. We discuss the possible origin of this novel superconducting state.

\section{Experimental Methods}
Single crystals of Ta$_4$Pd$_3$Te$_{16}$ were grown by a self-flux method \cite{TPT-Jacs}. The shiny crystals in flattened needle shape have the longest dimension along $b$ axis (the chain direction). The largest natural surface with typical dimensions of 2.5 $\times$ 0.25 mm$^2$ is in ($\overline{1}$03) plane, which is the $a'b$ plane in Fig. \ref{Crystal}(a). The thickness along the $c^*$ direction is about 0.1 mm. The dc magnetization measurements were performed in a SQUID (MPMS, Quantum Design), with an applied field of $H = 10$ Oe parallel to the $b$ direction. Four contacts were made directly on the sample surfaces with silver paint, which were used for both resistivity and thermal conductivity measurements along the $b$ direction at ambient pressure. The resistivity was measured in a $^4$He cryostat from 300 K to 2 K, and in a $^3$He cryostat down to 0.3 K. The thermal conductivity was measured in a dilution refrigerator, using a standard four-wire steady-state method with two RuO$_2$ chip thermometers, calibrated {\it in situ} against a reference RuO$_2$ thermometer. For resistivity measurements under pressure, the contacts were made with silver epoxy. Samples were pressurized in a piston-cylinder clamp cell made of Be-Cu alloy, with Daphne oil as the pressure media. The pressure inside the cell was determined from the $T_c$ of a tin wire.

\begin{figure} [t]
\centering
\includegraphics[clip,width=6cm]{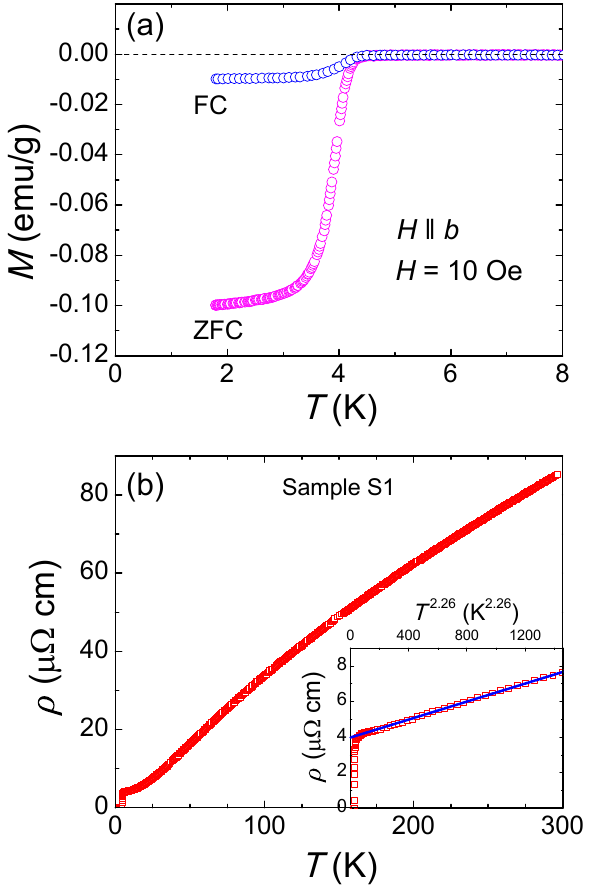}
\caption{\label{Resistivity}
(a) The dc magnetization at $H = 10$ Oe for Ta$_4$Pd$_3$Te$_{16}$ single crystal, with both zero-field-cooling (ZFC) and field-cooling (FC) processes. (b) The resistivity $\rho(T)$ along the $b$ direction of Ta$_4$Pd$_3$Te$_{16}$ single crystal (Sample S1) in zero field. The data between 7 and 25 K can be well fitted to $\rho(T)$ = $\rho_0$ + $AT^n$, giving residual resistivity $\rho_0$ = 3.96 $\mu \Omega$~cm and $n$ = 2.26, as shown in the inset.
}
\end{figure}

\section{Results and discussions}
Figure~\ref{Resistivity}(a) shows the typical low-temperature dc magnetization of Ta$_4$Pd$_3$Te$_{16}$ single crystal. With zero-field-cooling process, a sharp diamagnetic superconducting transition is observed at $T_c\approx4.3$ K. In Fig.~\ref{Resistivity}(b), the resistivity of Ta$_4$Pd$_3$Te$_{16}$ single crystal (Sample S1) in zero field is plotted. The resistivity decreases smoothly from room temperature to $T_c$. Fitting the data between 7 and 25 K to $\rho(T)$ = $\rho_0$ + $AT^n$ gives residual resistivity $\rho_0=3.96$ $\mu \Omega$~cm and $n$ = 2.26. The $T_c$ = 4.3 K is defined by $\rho$ = 0, which agrees well with the magnetization measurement.

\begin{figure} [t]
\centering
\includegraphics[clip,width=6cm]{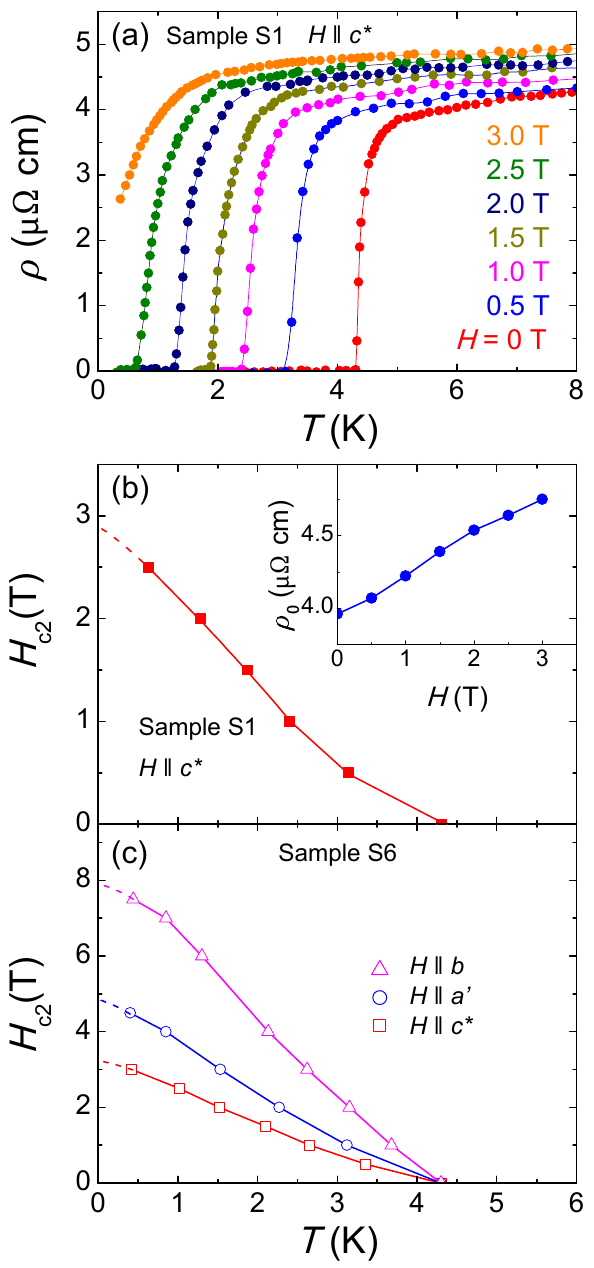}
\caption{\label{Fieldresistivity}
(a) Low-temperature resistivity of Ta$_4$Pd$_3$Te$_{16}$ single crystal (Sample S1) in magnetic field $H \parallel c^*$ up to 3 T. (b) The upper critical field $H_{c2}$ of Sample S1, defined by $\rho=0$. The dashed line is a guide to the eye, which points to $H_{c2}(0) \approx 2.9$ T. The inset shows the field dependence of $\rho_0$. (c) The $H_{c2}$ of Ta$_4$Pd$_3$Te$_{16}$ single crystal (Sample S6) for $H \parallel b$, $a'$, and $c^*$. The extrapolated $H_{c2}$ values at zero temperature are 7.9, 4.8, and 3.2 T, respectively.
}
\end{figure}

To determine the upper critical field $H_{c2}$(0) of Ta$_4$Pd$_3$Te$_{16}$, we measure the resistivity of Sample S1 down to 0.3 K in various magnetic fields along $c^*$ direction. Figure ~\ref{Fieldresistivity}(a) shows the low-temperature resistivity in fields up to 3 T. The temperature dependence of $H_{c2}$, defined by $\rho=0$ on the resistivity curves in Fig.~\ref{Fieldresistivity}(a), is plotted in Fig.~\ref{Fieldresistivity}(b). The dashed line is a guide to the eye, which points to $H_{c2}(0)\approx2.9$ T. The inset shows the field dependence of $\rho_0$, which manifests positive magnetoresistance, with $\rho_0$(2T) = 4.54 $\mu \Omega$~cm.

The anisotropy of $H_{c2}$ along $b$, $a'$, and $c^*$ directions for Sample S6 is shown in Fig.~\ref{Fieldresistivity}(c). The resistivity data of Sample S6 are not shown, and the $H_{c2}$ is also defined by $\rho=0$. Along three directions, $H_{c2}(0) \approx$ 7.9, 4.8, and 3.2 T are estimated from Fig.~\ref{Fieldresistivity}(c). The initial $H_{c2}$ slopes are -1.61, -0.85, and -0.53 T/K, which corresponds to the $H_{c2}$ ratio of $3.0:1.6:1$ near $T_c$ for $H \parallel b:a':c^*$. According to anisotropic Ginzburg-Landau (GL) theory $H_{c2}^i/H_{c2}^j = \sqrt{\rho_j}/\sqrt{\rho_i}$ \cite{Hussey}, the resistivity ratio $\rho_{c^*}:\rho_{a'}:\rho_b \approx 9.0:3.5:1$ is roughly estimated. This anisotropy is consistent with the quasi-1D structure of layered Ta$_4$Pd$_3$Te$_{16}$. Note that this ratio is much smaller than those of quasi-1D superconductors LiMo$_6$O$_{17}$ and (TMTSF)$_2$PF$_6$ \cite{Hussey,IJLee,Mihaly}.

\begin{figure} [t]
\centering
\includegraphics[clip,width=6cm]{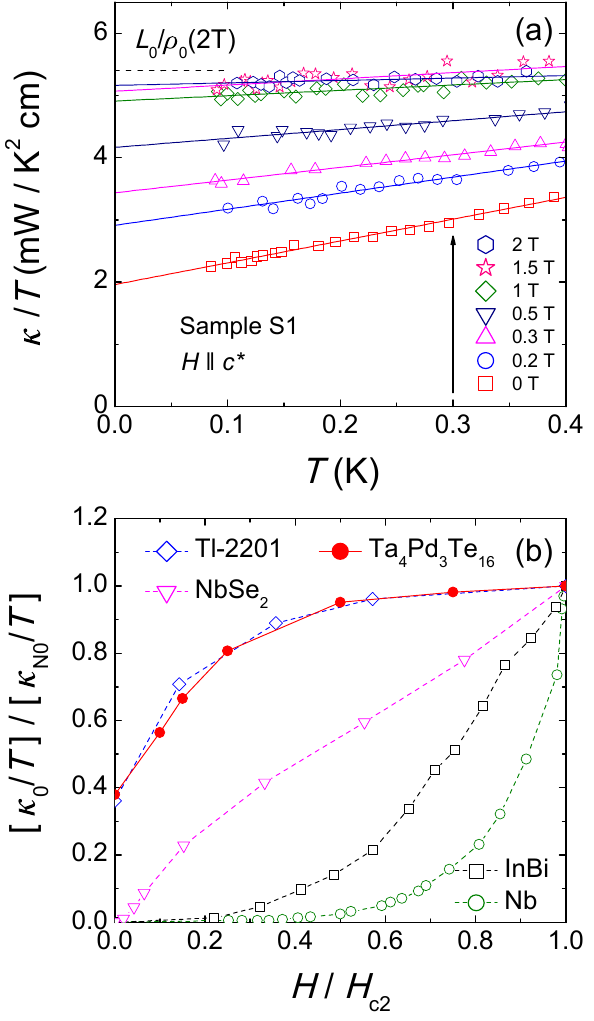}
\caption{\label{Thermal conductivity}
(a) Low-temperature thermal conductivity of Ta$_4$Pd$_3$Te$_{16}$ (Sample S1) in magnetic fields up to 2 T, applied along the $c^*$ direction. All the curves are roughly linear. The solid lines are fits to $\kappa/T$ = $a + bT$. The dashed line is the normal-state Wiedemann-Franz law expectation $L_0/\rho_0$(2T), where $L_0$ is the Lorenz number 2.45 $\times$ 10$^{-8}$ W $\Omega$ K$^{-2}$ and $\rho_0$(2T) = 4.54 $\mu \Omega$~cm. (b) Normalized $\kappa_0/T$ of Ta$_4$Pd$_3$Te$_{16}$ as a function of $H/H_{c2}$. Similar data of the clean $s$-wave superconductor Nb \cite{Lowell}, the dirty $s$-wave superconducting alloy InBi \cite{Willis}, the multiband $s$-wave superconductor NbSe$_2$ \cite{Boaknin}, and an overdoped $d$-wave superconductor Tl-2201 \cite{Proust} are also shown for comparison. The normalized $\kappa_0(H)/T$ of Ta$_4$Pd$_3$Te$_{16}$ clearly mimics that of Tl-2201.
}
\end{figure}

Low-temperature heat transport is an established bulk technique to probe the superconducting gap structure \cite{Shakeripour}. The thermal conductivity results of Ta$_4$Pd$_3$Te$_{16}$ single crystal (Sample S1) are presented in Fig.~\ref{Thermal conductivity}. Figure~\ref{Thermal conductivity}(a) shows the temperature dependence of thermal conductivity in magnetic field $H \parallel c^*$ up to 2 T, plotted as $\kappa/T$ vs $T$. The thermal conductivity at very low temperature can be usually fitted to $\kappa/T$ = $a$ + $bT^{\alpha-1}$ \cite{Mike,SYLi1}, in which the two terms $aT$ and $bT^\alpha$ represent contributions from electrons and phonons, respectively. The power $\alpha$ is typically between 2 and 3, due to specular reflections of phonons at the boundary \cite{Mike,SYLi1}. Since all the curves in Fig.~\ref{Thermal conductivity}(a) are roughly linear, we fix $\alpha$ to 2. The low values of $\alpha$ have been previously observed in several superconductors, for example, Cu$_{0.06}$TiSe$_2$ ($\alpha \approx$ 2.27) and KFe$_2$As$_2$ ($\alpha \approx$ 2) \cite{SYLi2,KFeAs}. Here, we only focus on the electronic term.

\begin{figure} [t]
\centering
\includegraphics[clip,width=8.5cm]{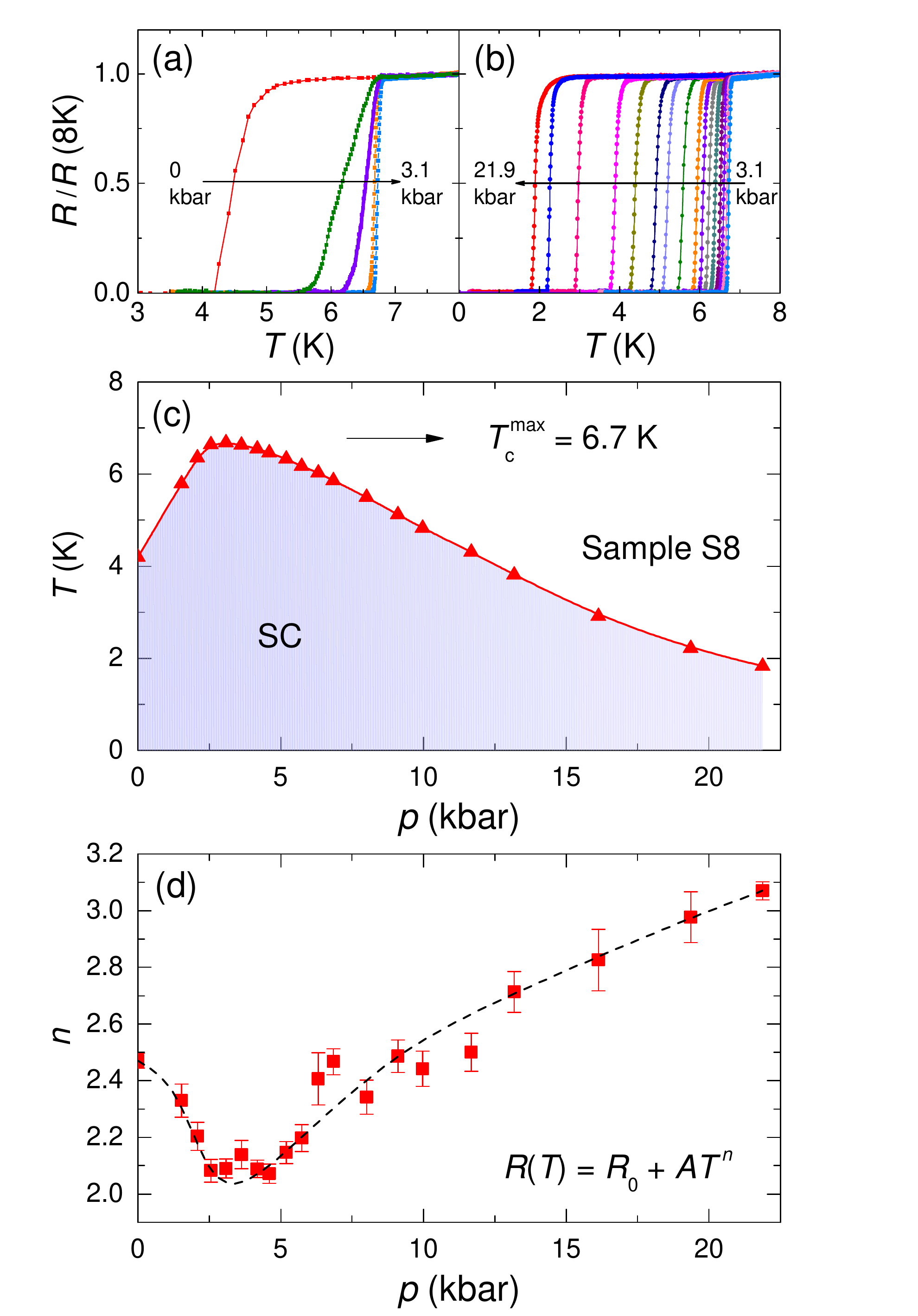}
\caption{\label{Pressureresistivity}
(a) and (b) Low-temperature resistance of Ta$_4$Pd$_3$Te$_{16}$ single crystal (Sample S8) under various pressures up to 21.9 kbar. (c) The pressure dependent $T_c$, defined by $\rho= 0$. There is a clear superconducting dome, with maximum $T_c$ = 6.7 K at optimal pressure $p_c$ = 3.1 kbar. (d) The pressure dependence of the exponent $n$ of resistance. There is a clear minimum of $n$ near $p_c$.}
\end{figure}

In zero field, the fitting gives a residual linear term with coefficient $\kappa_0/T \equiv a = 1.96 \pm 0.02$ mW K$^{-2}$~cm$^{-1}$. This value is more than 30$\%$ of the normal-state Wiedemann-Franz law expectation $\kappa_{N0}/T = L_0/\rho_0$(0T) = 6.19 mW K$^{-2}$~cm$^{-1}$, where $L_0 =$ 2.45  $\times$ 10$^{-8}$ W $\Omega$ K$^{-2}$ is the Lorenz number and $\rho_0$(0T) = 3.96 $\mu \Omega$~cm. In nodeless superconductors, all electrons become Cooper pairs as $T\rightarrow0$ and there are no fermionic quasiparticles to conduct heat. Therefore there is no residual linear term of $\kappa$, i.e., $\kappa_0/T = 0$. However, for unconventional superconductors with nodes in the superconducting gap, the nodal quasiparticles will contribute a finite $\kappa_0/T$ in zero field \cite{Shakeripour}. For example, $\kappa_0/T$ = 1.41 mW K$^{-2}$~cm$^{-1}$ for the overdoped $d$-wave cuprate superconductor Tl$_2$Ba$_2$CuO$_{6+\delta}$ (Tl-2201), which is about 36$\%$ of its $\kappa_{N0}/T$ \cite{Proust}, and $\kappa_0/T$ = 17 mW K$^{-2}$~cm$^{-1}$ for the $p$-wave superconductor Sr$_2$RuO$_4$, which is about 9$\%$ of its $\kappa_{N0}/T$ \cite{Suzuki}. The significant $\kappa_0/T$ ($>30\%\ \kappa_{N0}/T$) of Ta$_4$Pd$_3$Te$_{16}$ in zero field rules out the case that it results from a small nonsuperconducting metallic portion in the sample, thus it is a strong evidence for the presence of nodes in the superconducting gap \cite{Shakeripour}.


From Fig.~\ref{Thermal conductivity}(a), a small field $H = 0.2$ T has significantly increased the $\kappa/T$. Above $H = 1$ T, $\kappa/T$ tends to saturate. For $H = 1.5$ and 2 T, $\kappa_0/T = 5.07 \pm 0.05$ and 5.16 $\pm$ 0.04 mW K$^{-2}$~cm$^{-1}$ were obtained from the fittings, respectively. The value of $\kappa_0/T$ for $H = 2$ T roughly meets the normal-state Wiedemann-Franz law expectation $L_0/\rho_0$(2T) = 5.40 mW K$^{-2}$~cm$^{-1}$, which validates our method of extrapolating to $T \to 0$. We take $H=2$ T as the bulk $H_{c2}$(0) of Ta$_4$Pd$_3$Te$_{16}$. To choose a slightly different bulk $H_{c2}$(0) does not affect our discussions of the field dependence of $\kappa_0/T$ below.

In Fig.~\ref{Thermal conductivity}(b), the normalized $\kappa_0/T$ of Ta$_4$Pd$_3$Te$_{16}$ is plotted as a function of $H/H_{c2}$, together with the clean $s$-wave superconductor Nb \cite{Lowell}, the dirty $s$-wave superconducting alloy InBi \cite{Willis}, the multiband $s$-wave superconductor NbSe$_2$ \cite{Boaknin}, and the overdoped $d$-wave cuprate superconductor Tl-2201 \cite{Proust}. For Ta$_4$Pd$_3$Te$_{16}$, the field dependence of $\kappa_0/T$ clearly mimics the behaviour of Tl-2201. This rapid increase of $\kappa_0/T$ in magnetic field further rules out the case that the significant $\kappa_0/T$ results from a small nonsuperconducting metallic portion in the sample, since it should not change so dramatically in magnetic field. The rapid increase of $\kappa_0/T$ in magnetic field should come from the Volovik effect of nodal quasiparticles, thus provides further evidence for nodes in the superconducting gap \cite{Shakeripour}. To our knowledge, so far all nodal superconductors have unconventional pairing mechanism \cite{Norman-review}. In this regard, the nodal gap we demonstrate from thermal conductivity results suggests unconventional superconductivity in Ta$_4$Pd$_3$Te$_{16}$.


To get further clue to the pairing mechanism in Ta$_4$Pd$_3$Te$_{16}$, we map out its temperature-pressure phase diagram by resistivity measurement under pressures. Figure ~\ref{Pressureresistivity}(a) and ~\ref{Pressureresistivity}(b) present the low-temperature resistivity of Ta$_4$Pd$_3$Te$_{16}$ single crystal (Sample S8) under various pressures up to 21.9 kbar. At ambient pressure, the $T_c$ is 4.2 K, defined by $\rho = 0$. With increasing pressure, the $T_c$ first increases sharply to 6.7 K at 3.1 kbar, enhanced by 60$\%$. Then it decreases slowly to 1.8 K at 21.9 kbar. The non-monotonic pressure dependence of $T_c$ is plotted in Fig.~\ref{Pressureresistivity}(c), which shows a clear superconducting dome.

A temperature-pressure ($T_c$ vs $p$) or temperature-doping ($T_c$ vs $x$) superconducting dome has been commonly observed in many unconventional superconductors, including heavy-fermion superconductors, cuprate superconductors, iron-based superconductors, and Q2D organic superconductors \cite{Norman-review}. For example, the heavy-fermion superconductor CeCoIn$_5$ manifests a $T_c$ vs $p$ superconducting dome, and the unconventional superconductivity may result from the antiferromagnetic spin fluctuations \cite{Sidorov}. Theoretically, it has been shown that unconventional superconductivity can also appear on the boarder of a density transition, and the superconductivity is mediated by density fluctuations \cite{Monthoux1,Monthoux2}. This may be the case of the pressure-induced superconductivity in 1$T$-TiSe$_2$, with the superconducting dome appearing around the critical pressure related to the charge-density wave (CDW) meltdown \cite{Kusmartseva}.

For Ta$_4$Pd$_3$Te$_{16}$, recent electronic structure calculations showed that its electronic states are mostly derived from Te $p$ states with small Ta $d$ and Pd $d$ contributions, which places the compound far from magnetic instabilities \cite{Singh}. Two scanning tunneling microscopy (STM) studies on Ta$_4$Pd$_3$Te$_{16}$ found commensurate modulations along atom chains, which may arise from CDW \cite{Du,QFan}. CDW usually appears in low-dimensional compounds, such as TTF-TCNQ, NbSe$_3$, and NbSe$_2$ \cite{Andrieux,Briggs,HNSLee}, therefore it is not surprising that CDW exists in layered Ta$_4$Pd$_3$Te$_{16}$ with Q1D structure. The absence of resistivity anomaly in the resistivity curve suggests that the CDW in Ta$_4$Pd$_3$Te$_{16}$ is quite weak, since the robustness of CDW can be reflected from the resistivity anomaly, as seen in TTF-TCNQ, NbSe$_3$, and NbSe$_2$ \cite{Andrieux,Briggs,HNSLee}.

To examine whether the superconducting dome relates to a CDW meltdown in Ta$_4$Pd$_3$Te$_{16}$, we carefully fit the resistance data up to 25 K to $\rho(T)$ = $\rho_0$ + $AT^n$ for each pressure, and plot the pressure dependence of the exponent $n$ in Fig. 5(d). There is a clear minimum of $n$ near the optimal pressure $p_c$ = 3.1 kbar. Similar pressure dependence of $n$ has been observed in 1$T$-TiSe$_2$ \cite{Kusmartseva}. For 1$T$-TiSe$_2$, the $n$ = 3 at pressure above 40 kbar is attributed to the phonon-assisted $s-d$ interband scattering, and the suppression of $n$ in the 20 - 40 kbar pressure region signifies the presence of CDW fluctuations around a critical pressure of 30 kbar \cite{Kusmartseva}. In this context, $p_c$ = 3.1 kbar is likely the critical pressure where the CDW in Ta$_4$Pd$_3$Te$_{16}$ is completely suppressed. If this is the case, the nodal superconductivity in Ta$_4$Pd$_3$Te$_{16}$ may originate from the CDW fluctuations.

\section{Conclusion}
We study the superconducting gap structure of new layered superconductor Ta$_4$Pd$_3$Te$_{16}$ by low-temperature thermal conductivity measurements. The significant $\kappa_0/T$ in zero magnetic field and its rapid field dependence suggest nodes in the superconducting gap. Further measurements of resistivity under pressure reveal a superconducting dome in the temperature-pressure phase diagram. These results indicate unconventional superconductivity in Ta$_4$Pd$_3$Te$_{16}$. With the recent STM evidence for the existence of CDW and our observation of the suppression of exponent $n$ near the optimal pressure $p_c$, Ta$_4$Pd$_3$Te$_{16}$ may provide a rare platform to study the unconventional superconductivity near a CDW instability. Clarifying the pairing symmetry and mechanism of this new layered superconductor will give us new understandings of unconventional superconductivity.

\begin{acknowledgments}
We thank X. H. Chen, J. K. Dong, and D. L. Feng for helpful discussions. This work is supported by the Natural Science Foundation of China, the Ministry of Science and Technology of China (National Basic Research Program No. 2012CB821402 and 2015CB921401), Program for Professor of Special Appointment (Eastern Scholar) at Shanghai Institutions of Higher Learning, and STCSM of China (No. 15XD1500200).
\end{acknowledgments}

\nocite{*}


\end{document}